\documentclass[12]{article}
\usepackage{fleqn}
\usepackage{graphicx}
\begin{document}
\Large
\begin{center}
{\bf Space versus Time:\\ Unimodular versus Non-Unimodular\\
Projective Ring Geometries?}
\end{center}
\large
\vspace*{-.1cm}
\begin{center}
Metod Saniga$^{1}$
and
Petr Pracna$^{2}$

\end{center}
\vspace*{-.4cm} \normalsize
\begin{center}
$^{1}$Astronomical Institute, Slovak Academy of Sciences\\
SK-05960 Tatransk\' a Lomnica, Slovak Republic\\
(msaniga@astro.sk)

\vspace*{.0cm} and

\vspace*{.0cm}

$^{2}$J. Heyrovsk\' y Institute of Physical
Chemistry, v.v.i., Academy of Sciences of \\ the Czech Republic, Dolej\v
skova 3, CZ-182 23 Prague 8, Czech
Republic\\
(pracna@jh-inst.cas.cz)

\end{center}

\vspace*{-.3cm} \noindent \hrulefill

\vspace*{.0cm} \noindent {\bf Abstract}

\noindent Finite projective (lattice) geometries defined over
rings instead of fields have recently been recognized to be of
great importance for quantum information theory. We believe that
there is much more potential hidden in these geometries to be
unleashed for physics. There exist specific rings over which the
projective spaces feature two principally distinct kinds of basic
constituents (points and/or higher-rank linear subspaces), intricately
interwoven with each other --- unimodular and non-unimodular. We
conjecture that these two projective ``degrees of freedom" can
rudimentary be associated with spatial and temporal dimensions of
physics, respectively. Our hypothesis is illustrated on the
projective line over the smallest ring of ternions. Both the
fundamental difference and intricate connection between time and
space are demonstrated, and even the ring geometrical germs of the
observed macroscopic dimensionality (3+1) of space-time and the arrow of time are
outlined. Some other conceptual implications of this speculative
model (like a hierarchical structure of physical systems) are also mentioned.

\vspace*{.2cm}

\vspace*{-.2cm} \noindent \hrulefill

\vspace*{-.2cm} \noindent
\section{Introduction}

\vspace*{-.3cm} If a theoretical physicist working on unification
of quantum mechanics and general relativity is asked to pinpoint
the most serious problems they face, the answer will most likely
be: the interpretation of the {\it time} and the observed
macroscopic {\it dimensionality of space-time}. The first problem
of the two is central to any approach to quantum gravity that
ascribes, by first instance, an important role to classical
general relativity and stems from the fundamentally different
roles played by the concept of time in quantum theory and in
general relativity. In quantum theory, time does not behave as a
physical quantity in the usual sense, since --- unlike spatial
coordinates
--- it is not represented by an operator, being rather treated as a
background parameter employed to mark the evolution of the system. Moreover, the notion of
an event happening at a given time plays a crucial role in the technical and conceptual
foundations of quantum theory. Classical general relativity, however, handles time in a
very different manner. Time is not regarded as a background parameter, even in the broad-minded
sense of special relativity, namely as an aspect of a fixed, background space-time structure.
It is rather our {\it interpretation} of the concept of time which makes us to either view our
reality as evolving in three dimensions (``arrow"of time), or as being some kind of a four-dimensional
construct (``frozen" time).

Typically, physicists simply take the observed macroscopic
dimensionality of space-time for granted and do not bother why it
has just four rather than any other number of dimensions. Only
seldom do they embark upon speculations in this respect. Such
contemplations begin, as a rule, by noticing that a given physical
theory has considerably different properties in space-times of
different dimensionality. By claiming that some feature/s of the
theory typical of the four-dimensional space-time is/are
fundamental, a reason can be offered for our space-time having the
observed number of dimensions. These explorations have, apart from the
quest for a rational explanation of the dimensionality of the
physical space-time, also a more practical aspect. Quantum field
theories are notorious for being ill defined in four-dimensional
space-time but can often be seen to fare well in space-times of a
different dimensionality. Successes of quantum field theory in
both higher- and lower-dimensional space-times almost lead to an
ironical statement that the only reason for our universe to be
endowed with four dimensions is the irrationality of such a
choice: the four-dimensional space-time seems the most problematic
setting for a quantum field theory to work. Among further
physically-based arguments, it is worth mentioning Weyl's
well-known observation that the Maxwell equations are tied
uniquely to the $3 + 1$ space-time, and/or intriguing Ehrenfest's
reasoning that stable atoms are only possible in $3 + 1$
dimensions. Another class of well-known heuristic inquiries is
more mathematically oriented. Here we can, for example, rank the
fact that the Weyl tensor, which in Einstein's gravitation theory
carries information about that part of the space-time curvature
which is not locally determined by the energy-momentum, vanishes
in less than four dimensions, or a topological reason that $n \neq
4$ dimensional manifolds always feature a unique differentiable
structure, while those with $n = 4$ do not. Finally, there is a
large and still growing group of scholars who favour the so-called
anthropic principle for a rational explanation of the
macro-dimensionality of space-time.

It may well be that the two above-mentioned problems of quantum gravity, viz.
the ``strange behaviour" of time and the fact that our Universe features just
four macroscopic dimensions, are intimately linked with each other, being in
fact the two sides of the same coin. This is also the point of view adopted in
this paper. In what follows we shall introduce a simple mathematical model
which gives a sound formal footing to such a hypothesis. The model rests on the
concept of the projective line defined over a ring instead of a field (Blunck
and Havlicek 2000; Veldkamp 1995; Herzer 1995; and Blunck and Herzer 2005); the
principal difference between the two kinds of geometry lies in the fact that
whereas in a field every non-zero element has its inverse, in a ring which is
not a field there exist non-zero elements lacking inverses (and so in this case
we cannot introduce the operation of division, i.\,e., multiplication by the
inverses). Although this concept was introduced into physics only recently, it
has already produced a number of crucial insights into the nature of
finite-dimensional quantum systems (see, e.\,g., Havlicek and Saniga 2008a;
Saniga, Planat and Pracna 2008; Planat and Baboin 2007, and references
therein). The model proposed, and its envisaged higher-order generalizations,
will serve as another illustration of our belief that there is much potential
hidden in these remarkable finite geometries to be unleashed for physics.

\section{Smallest Line over Ternions --- the Seed of Space-Time?}

\vspace*{-.3cm}
We shall consider a finite associative ring with unity $1(\neq 0)$, $R$, and denote the
{\it left} module on two generators over $R$  by $R^2$. The set
$R(r_1, r_2)$, defined as follows
\begin{equation}
R(r_1, r_2):= \left\{ (\alpha r_1, \alpha r_2) | (r_1, r_2) \in R^2, \alpha \in R \right\},
\end{equation}
is a left {\it cyclic} submodule of $ R^2$. Any such submodule
is called {\it free} if the mapping $\alpha \mapsto (\alpha r_1,
\alpha r_2) $ is injective, i.\,e., if
$(\alpha r_1, \alpha r_2) $ are all {\it
distinct}. Next, we shall call $(r_1, r_2) \in
R^2$ {\it uni}modular if there exist elements $x_1$ and
$x_2$ in $R$ such that
\begin{equation}
r_1 x_1 + r_2 x_2 = 1.
\end{equation}
It can easily be shown that if $(r_1, r_2)$ is unimodular, then $R(r_1, r_2)$
is free; any such free cyclic submodule represents a point of the projective
line defined over $R$, $P(R)$ (Blunck and Havlicek 2000; Veldkamp 1995; Herzer
1995; and Blunck and Herzer 2005):
\begin{equation}
P(R) := \left\{ R(r_1, r_2)| (r_1, r_2)~ {\rm unimodular} \right\}.
\end{equation}
We just mention in passing that in any such geometry a point is a {\it set} of
pairs/vectors (of cardinality $|R|$) and has thus a subtle internal structure,
which is in sharp contrast with Euclid's point of view that ``a point is that
which has no parts." Obviously, every projective line over any ring features
free cyclic submodules generated by unimodular vectors (in the sequel also
called unimodular points) and over a vast majority of finite rings these are
{\it the only} free cyclic submodules of $R^2$. Yet, as we shall soon see,
there are also rings which in addition yield free cyclic submodules generated
by {\it non}-unimodular vectors. In light of this fact, and following also the
spirit and strategy of Brehm, Greferath and Schmidt (1995), it is reasonable to
consider a more general concept of the projective ring line, namely
\begin{equation}
\widehat{P}(R) := \left\{ R(r_1, r_2)| R(r_1, r_2)~ {\rm free} \right\}.
\end{equation}
So
\begin{equation}
\widehat{P}(R) = P(R) \cup \widetilde{P}(R)
\end{equation}
with $\widetilde{P}(R)$ standing for the part of the projective line comprising
solely the points generated by non-unimodular vectors (also referred to as
non-unimodular points). The two parts of this generalized ring line, if both
non-empty, are, on the one side, very different from each other, yet, on the
other side, intricately interwoven with each other. And it is this relation
between the two parts of the line which is central to our subsequent
discussions.

In order to see this explicitly, we shall have a detailed look at
such generalized projective line over the smallest ring of
ternions $R_{\diamondsuit}$, i.\,e. the ring defined as follows
\begin{equation}
R_{\diamondsuit}  \equiv \left\{ \left(
\begin{array}{cc}
a & b \\
0 & c \\
\end{array}
\right) \mid ~ a, b, c \in GF(2) \right\},
\end{equation}
where $GF(2)$ is the Galois field of two elements. This is, up to isomorphisms,
the unique non-commutative ring of order eight, endowed with two invertible
elements (units) and six zero-divisors, and, most interestingly, also the
smallest ring where $\widetilde{P}(R)$ is not an empty set (Saniga, Havlicek,
Planat and Pracna 2008; Havlicek and Saniga 2008b). Employing its addition and
multiplication tables (Table 1), we readily find altogether 36 {\it uni}modular
vectors which generate 18 different free cyclic submodules, namely
\begin{table}[t]
\begin{center}
\caption{Addition ({\it left}) and multiplication ({\it right}) in
$R_{\diamondsuit}$.} \vspace*{0.2cm}
\begin{tabular}{||c|cccccccc||}
\hline \hline
$+$ & 0 & 1 & 2 & 3 & 4 & 5 & 6 & 7 \\
\hline
0 & 0 & 1 & 2 & 3 & 4 & 5 & 6 & 7 \\
1 & 1 & 0 & 6 & 7 & 5 & 4 & 2 & 3 \\
2 & 2 & 6 & 0 & 4 & 3 & 7 & 1 & 5 \\
3 & 3 & 7 & 4 & 0 & 2 & 6 & 5 & 1 \\
4 & 4 & 5 & 3 & 2 & 0 & 1 & 7 & 6 \\
5 & 5 & 4 & 7 & 6 & 1 & 0 & 3 & 2 \\
6 & 6 & 2 & 1 & 5 & 7 & 3 & 0 & 4 \\
7 & 7 & 3 & 5 & 1 & 6 & 2 & 4 & 0 \\
\hline \hline
\end{tabular}~~~~~
\begin{tabular}{||c|cccccccc||}
\hline \hline
$\times$ & 0 & 1 & 2 & 3 & 4 & 5 & 6 & 7  \\
\hline
0 &  0 & 0 & 0 & 0 & 0 & 0 & 0 & 0 \\
1 &  0 & 1 & 2 & 3 & 4 & 5 & 6 & 7 \\
2 &  0 & 2 & 1 & 3 & 7 & 5 & 6 & 4 \\
3 &  0 & 3 & 5 & 3 & 6 & 5 & 6 & 0 \\
4 &  0 & 4 & 4 & 0 & 4 & 0 & 0 & 4 \\
5 &  0 & 5 & 3 & 3 & 0 & 5 & 6 & 6 \\
6 &  0 & 6 & 6 & 0 & 6 & 0 & 0 & 6 \\
7 &  0 & 7 & 7 & 0 & 7 & 0 & 0 & 7 \\
\hline \hline
\end{tabular}
\end{center}
\end{table}

\vspace*{.2cm}
~~~$R_{\diamondsuit}(1,0)=R_{\diamondsuit}(2,0) = \left\{(0,0), (6,0), (4,0), (7,0), (5,0), (3,0), (2,0), (1,0) \right\}$,

~~~$R_{\diamondsuit}(1,6)=R_{\diamondsuit}(2,6) = \left\{(0,0), (6,0), (4,0), (7,0), (5,6), (3,6), (2,6), (1,6) \right\}$,

~~~$R_{\diamondsuit}(1,3)=R_{\diamondsuit}(2,3) = \left\{(0,0), (6,0), (4,0), (7,0), (5,3), (3,3), (2,3), (1,3) \right\}$,

~~~$R_{\diamondsuit}(1,5)=R_{\diamondsuit}(2,5) = \left\{(0,0), (6,0), (4,0), (7,0), (5,5), (3,5), (2,5), (1,5) \right\}$,

~~~$R_{\diamondsuit}(7,3)=R_{\diamondsuit}(4,3) = \left\{(0,0), (6,0), (4,0), (7,0), (0,3), (6,3), (4,3), (7,3) \right\}$,

~~~$R_{\diamondsuit}(7,5)=R_{\diamondsuit}(4,5) = \left\{(0,0), (6,0), (4,0), (7,0), (0,5), (6,5), (4,5), (7,5) \right\}$,

~~~$R_{\diamondsuit}(1,7)=R_{\diamondsuit}(2,4) = \left\{(0,0), (6,6), (4,4), (7,7), (5,6), (3,0), (2,4), (1,7) \right\}$,

~~~$R_{\diamondsuit}(1,4)=R_{\diamondsuit}(2,7) = \left\{(0,0), (6,6), (4,4), (7,7), (5,0), (3,6), (2,7), (1,4) \right\}$,

~~~$R_{\diamondsuit}(1,1)=R_{\diamondsuit}(2,2) = \left\{(0,0), (6,6), (4,4), (7,7), (5,5), (3,3), (2,2), (1,1) \right\}$,

~~~$R_{\diamondsuit}(1,2)=R_{\diamondsuit}(2,1) = \left\{(0,0), (6,6), (4,4), (7,7), (5,3), (3,5), (2,1), (1,2) \right\}$,

~~~$R_{\diamondsuit}(4,1)=R_{\diamondsuit}(7,2) = \left\{(0,0), (6,6), (4,4), (7,7), (0,5), (6,3), (7,2), (4,1) \right\}$,

~~~$R_{\diamondsuit}(7,1)=R_{\diamondsuit}(4,2) = \left\{(0,0), (6,6), (4,4), (7,7), (0,3), (6,5), (4,2), (7,1) \right\}$,

~~~$R_{\diamondsuit}(3,7)=R_{\diamondsuit}(3,4) = \left\{(0,0), (0,6), (0,4), (0,7), (3,0), (3,6), (3,4), (3,7) \right\}$,

~~~$R_{\diamondsuit}(5,7)=R_{\diamondsuit}(5,4) = \left\{(0,0), (0,6), (0,4), (0,7), (5,0), (5,6), (5,4), (5,7) \right\}$,

~~~$R_{\diamondsuit}(5,1)=R_{\diamondsuit}(5,2) = \left\{(0,0), (0,6), (0,4), (0,7), (5,5), (5,3), (5,2), (5,1) \right\}$,

~~~$R_{\diamondsuit}(3,1)=R_{\diamondsuit}(3,2) = \left\{(0,0), (0,6), (0,4), (0,7), (3,5), (3,3), (3,2), (3,1) \right\}$,

~~~$R_{\diamondsuit}(6,1)=R_{\diamondsuit}(6,2) = \left\{(0,0), (0,6), (0,4), (0,7), (6,5), (6,3), (6,2), (6,1) \right\}$,

~~~$R_{\diamondsuit}(0,1)=R_{\diamondsuit}(0,2) = \left\{(0,0), (0,6), (0,4), (0,7), (0,5), (0,3), (0,2), (0,1) \right\}$,

\vspace*{.2cm}
\noindent
and six {\it non-}unimodular vectors giving rise to three distinct free cyclic submodules, namely

\vspace*{.2cm}
~~~$R_{\diamondsuit}(4,6)=R_{\diamondsuit}(7,6) = \left\{(0,0), (6,0), (0,6), (6,6), (4,0), (7,0), (7,6), (4,6) \right\}$,

~~~$R_{\diamondsuit}(4,7)=R_{\diamondsuit}(7,4) = \left\{(0,0), (6,0), (0,6), (6,6), (4,4), (7,7), (7,4), (4,7) \right\}$,

~~~$R_{\diamondsuit}(6,4)=R_{\diamondsuit}(6,7) = \left\{(0,0), (6,0), (0,6), (6,6), (0,4), (0,7), (6,7), (6,4) \right\}$.

\begin{figure}[pth!]
\centerline{\includegraphics[width=6.5truecm,clip=]{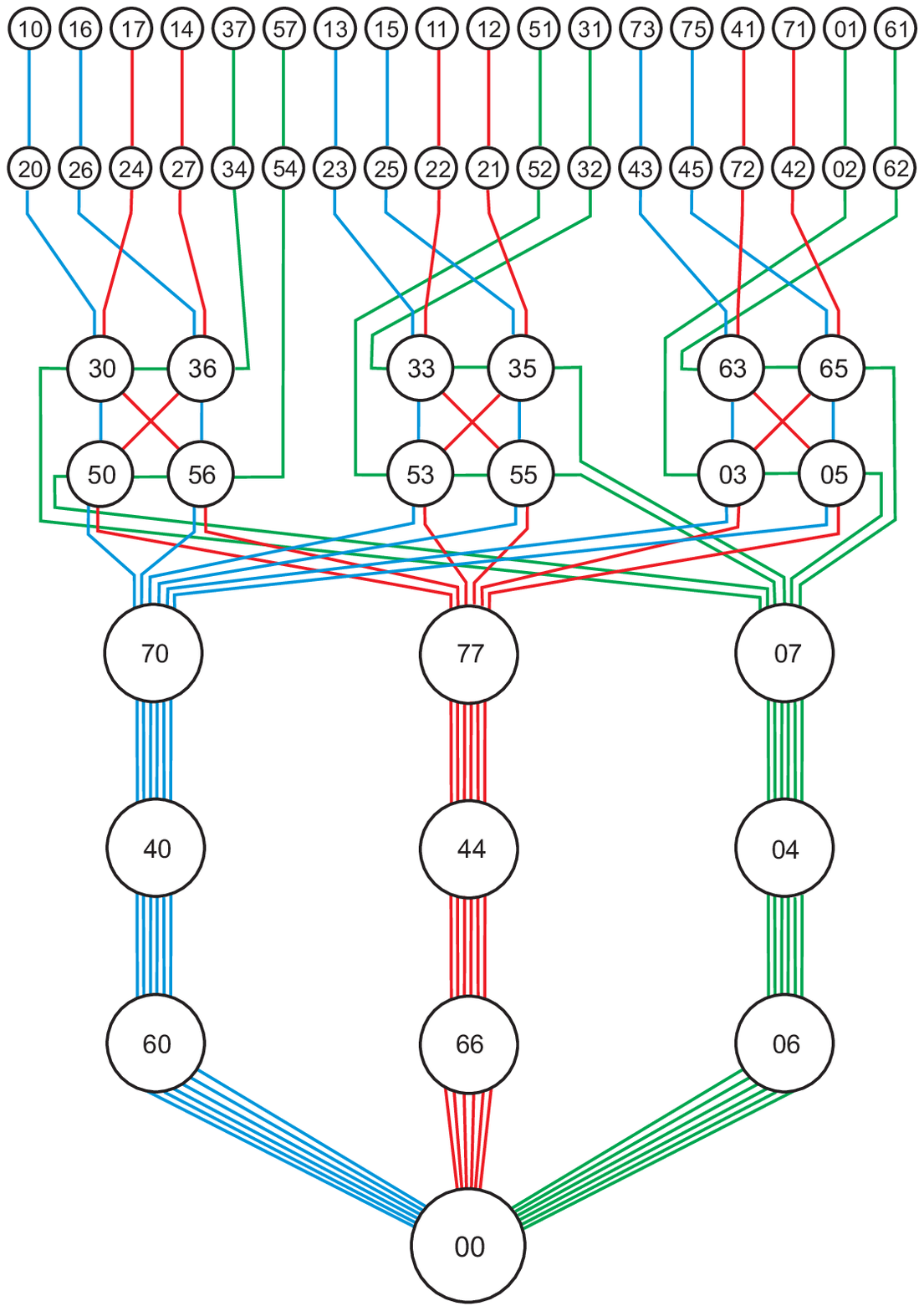}
\hspace*{1.5cm}
\includegraphics[width=2.23truecm,clip=]{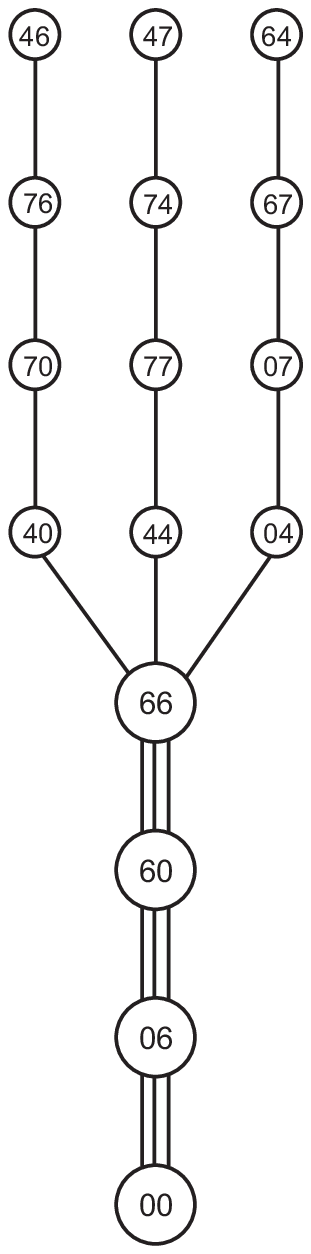}} \caption{A
diagrammatic illustration of the structure of the unimodular ({\it
left}) and non-unimodular ({\it right}) parts of the projective
line over the smallest ring of ternions. The symbols and notation
are explained in the text.}
\end{figure}

\begin{figure}[pth!]
\centerline{\includegraphics[width=7.6truecm,clip=]{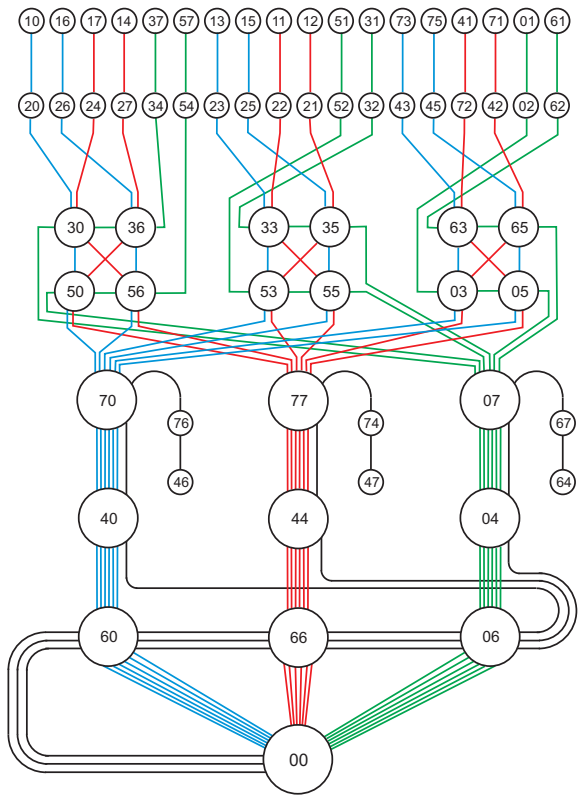}}
\caption{A diagrammatic sketch of the intricate link between the
two parts of the line shown in the preceding figure.}
\end{figure}

\vspace*{.2cm} \noindent The structure of both the sets and how
the two are intertwined can be fairly well visualised and grasped
in terms of a network of broken line-segments (polygons) as
depicted in Figures 1 and 2, respectively. In both figures, a
circle with an attached two-digit number $XY$ stands for the $(X,
Y)$ vector and line-joins of two circles indicate that the two
corresponding vectors lie on the same free cyclic submodule; the
size of vectors is roughly proportional to the number of
submodules they are contained in. Just a passing look at Figure 1 reveals
a principal distinction between the two ``sectors/regimes" of the line.

The first fact to be easily noticed is the difference in the cardinalities of
the two sets, a rather trivial issue. The second feature is a bit more
intricate: whilst in the unimodular configuration the only common element of
all the 18 points is the vector $(0,0)$, the three non-unimodular points share
(three) additional pairs. This latter property can be rigorously accounted for
after the concepts of {\it neighbour/distant} are introduced (Blunck and
Havlicek 2000; Veldkamp 1995; Herzer 1995; and Blunck and Herzer 2005). Given
the obvious fact that the $(0,0)$ vector belongs to every cyclic submodule, we
shall call two distinct points $R(r_1, r_2)$ and $R(s_1, s_2)$ of a projective
line distant if $|R(r_1, r_2) \cap R(s_1, s_2)| = 1$ and neighbour if $|R(r_1,
r_2) \cap R(s_1, s_2)| > 1$. We then find that {\it all} non-unimodular points
are pairwise {\it neighbour}, whereas the maximum number of mutually neighbour
points in the unimodular case is $6 \neq 18$ (any point is obviously neighbour
to itself). Hence, in the unimodular case it also makes sense to ask what the
maximum number of pairwise {\it distant} points is, the answer being
--- {\it three}. These facts are illustrated in Figure 1, left, by the use of three different colours. Let us
pick up one blue point, for example $R(1,0)$. Then all the points of the same
colour have the property of being pairwise neighbour. Hence, to find a distant
point to our selected blue point, we have to go to the sets of different
colour, say red. However, only four out of six red points are distant to our
selected point; we can pick up any of them, say $R(1,1)$. The last, third point
distant to the two selected must necessarily be a green one, and one can take
either $R(0,1)$ or $R(6,1)$. This reasoning can also be rephrased the other way
round; by choosing any triple (i.\,e., the maximum number) of pairwise distant
points, the set of unimodular points is naturally partitioned into (i.\,e.,
factored into a disjoint union of) three maximum sets of mutually neighbour
points, with different triples yielding {\it one and the same} partitioning
(the three sets distinguished by different colours in both Figure 1, left, and
Figure 2). The last pronounced difference between the two sectors is perhaps
most interesting and most intriguing as well. If we take any unimodular point,
we see that the only vectors that are unique to the point are its two
generators; that is, any other vector on each unimodular free cyclic submodule
belongs also to some other submodule(s)/point(s) (see Figure 1, left). If we
look at any of the three non-unimodular points (Figure 1, right), we find that
apart from its two generating vectors there are other two vectors that lie on
just this point. This ``peculiar" feature enables the so-called ``geometric
condensation" phenomenon to take place in terms of which the ``condensate" of
our non-unimodular sub-configuration is found to be isomorphic to nothing but
the ordinary projective line over $GF(2)$ (Saniga 2008); one simply associates
the set of the four common vectors $\{(0,0), (6,0), (0,6), (6,6)\}$ with the
$(0,0)$ vector and the remaining three quadruples $\{(4,0), (7,0), (7,6),
(4,6)\}$, $\{(4,4), (7,7), (7,4), (4,7)\}$ and $\{(0,4), (0,7), (6,7), (6,4)\}$
with the $(1,0)$, $(1,1)$ and $(0,1)$ vectors of the $GF(2)$-line,
respectively.\footnote{It is also worth mentioning here that it is precisely
this property that, at least in the case of rings of ternions, puts projective
lines on a different footing than any higher-dimensional projective spaces
(Havlicek and Saniga 2008b).} When it comes to the coupling between the two
parts of the line (Figure 2), another remarkable feature is encountered; {\it
every} non-unimodular point is {\it neighbour} to {\it every} unimodular one.
If we regard neighbourness between two points as a tighter link when compared
to the distant relation, we can also say that the coupling between every
non-unimodular and every unimodular point is more intimate than that between
any mutually distant (hence unimodular) points.

At this stage the amount of information gathered about the structure of our
smallest ternionic projective line is sufficient for us to make a step out of
pure mathematics and show that this line may be of great relevance for physics,
especially as a sort of conceptual guide for approaching the two major quantum
gravity problems discussed in the introduction. To this end in view, we have
first to realize that our line is rather simple as featuring only 21 points in
total, and the notions of neighbour/distant are purely algebraic ones (not the
slightest trace of metric whatsoever at this level!). Despite of these facts,
let us nevertheless make a daring hypothesis and take this configuration as a
finite prototype of space-time by tentatively identifying its unimodular part
with the ``seeds" of spatial degrees of freedom and its non-unimodular portion
with the ``buds" of time. In light of the above-discussed properties of the
line, such an identification immediately entails a crucial distinction between
space and time, the former being ``more heterogeneous and less compact"
(existence of both mutually neighbour and mutually distant unimodular points
compared with only mutually neighbour non-unimodular ones) and ``more complex"
(the unimodular set featuring six times more elements than the non-unimodular
one) than the latter. Moreover, given the unique partitioning of the unimodular
aggregate of points induced by any set of three pairwise distant members, our
ternionic spatial degrees of freedom are already endowed with something which
can be regarded as a first trace of the observed three-dimensionality of space;
each of the three maximum sets of mutually neighbour points viewed as the germ
of a single spatial dimension. In the same spirit, complete absence of the
notion of mutually distant on the non-unimodular set lends itself as a natural
explanation of the observed uni-dimensionality of time and the unique geometric
condensation phenomenon may well represent nothing but a ternionic ``germ" of
the arrow/unidirectionality of time. At this stage we cannot elaborate more on
how this condensation/contraction phenomenon gives rise to the microscopic
irreversibility that is required to occur in the laws of nature to explain the
arrow of time. But what is already firmly established is the impossibility of
making our time dimension two-directional, because that would require a sort of
natural ``expansion" of the non-unimodular part of our ternionic line to a
(part of) projective line over a certain bigger ring, which is clearly not the
case. Also, it is worth stressing that this arrow-of-time effect is strictly
due to a mathematical phenomenon.

This difference between space and time becomes even more pronounced after we
take into account the already-mentioned fact that rings generating both kinds
of points seem to occur rather sparsely when compared to those which do not
(only eight such rings out of 131 up to order 31, see below); that is, a
generic projective ring line favours the universe lacking the time dimension.

\section{What Is a Next Move?}

\vspace*{-.3cm} In order to make the above-described space-time model more
realistic, we obviously have to employ projective lines endowed with many more
points and, so, consider rings of much higher orders. When doing so, two
crucial, and rather severe, constraints have to be preserved: the line must be
endowed with both unimodular and non-unimodular points (to ``generate" both
space and time) and the maximum number of mutually distant (unimodular) points
must be three (to account for the observed dimensionality of space). To meet
the latter constraint the necessary condition is to focus only on rings of even
orders. The former constraint seems much more stringent and much more
problematic to ascertain because, to our best knowledge, not only is there no
general recipe for finding rings that yield both unimodular and non-unimodular
free cyclic submodules, but also very little is known about the corresponding
projective lines and/or higher-dimensional spaces (Brehm, Greferath and Schmidt
1995). We, therefore, decided for a case-by-case inspection of all the finite
rings up to order 31 (altogether 131, see also Saniga, Planat, Kibler and
Pracna 2007; Saniga, Planat and Pracna 2006) and found, apart
$R_{\diamondsuit}$, six non-isomorphic rings of order 16 (one of them being
isomorphic to $GF(2) \times R_{\diamondsuit}$) and a single ring of order 24
(that isomorphic to $GF(3) \times R_{\diamondsuit}$) to meet our constraints;
remarkably, all of them being, like $R_{\diamondsuit}$ itself, {\it
non}-commutative. The five rings of order 16 all feature 12 zero-divisors and
split into two distinct classes (A and B) differing from each other in the
cardinalities of their ideals of order 2, 4, and 8 --- 2, 2, 2 versus 3, 1, 2,
respectively. And although this difference does not manifest itself in the
unimodular sector of the associated projective lines (Saniga, Planat and Pracna
2006), in the non-unimodular sector it considerably does; for in the former
class this sector comprises six points and its condensate is isomorphic to
$P(Z_4)$ and/or $P(GF(2)[x]/\langle x^2 \rangle$), whilst in the latter class
it contains nine points and its condensed form is much more involved, not
having counterpart in any projective ring line (Saniga 2008). In light of our
hypothesis this would mean that we have two qualitatively distinct space-times
having identical spatial characteristics but substantially different time
dimensions. Moving to the line over $GF(2) \times R_{\diamondsuit}$ we find its
non-unimodular sector to condense into $P(GF(2) \times GF(2))$ and that of the
line over the ring of order 24 into $P(GF(2) \times GF(3))$. Our findings are,
for the reader's convenience, summarized in Table 2.

\begin{table}[ht]
\begin{center}
\caption{Basic properties of several small projective ring lines endowed with
both unimodular and non-unimodular sectors. The first column lists the type of the line, the second and third columns
show the cardinalities of its unimodular and non-unimodular sectors, respectively, and the last column
gives the character of the associated condensate.}
\vspace*{0.2cm}
\begin{tabular}{||l|c|c|l||}
\hline \hline
\vspace*{-.35cm}
&&&\\
$\widehat{P}(R_{\diamondsuit})$ & 18& 3& $P(GF(2))$\\
$\widehat{P}$(16/12A) & 36 & 6 & $P(Z_4)$ or $P(GF(2)[x]/\langle x^2 \rangle)$ \\
$\widehat{P}$(16/12B) & 36 & 9 & not a ring line(?) \\
$\widehat{P}(GF(2) \times R_{\diamondsuit})$ & 54 & 9 & $P(GF(2) \times GF(2))$\\
$\widehat{P}(GF(3) \times R_{\diamondsuit})$ & ~72~ & ~12~ & $P(GF(2) \times GF(3)) \simeq P(Z_6)$\\
\hline \hline
\end{tabular}
\end{center}
\end{table}

These few examples give us important clues as to what one can
expect when making use of higher-order ring lines. Remarkably, one
finds that, except for the case where the condensate is not a ring
line, the ratio between the number of unimodular and
non-unimodular points remains the same (six). When focusing on
condensates themselves, which practically contain all essential
information about the structure of the corresponding ``seed" time
dimensions, we see, on the one hand, a big discrepancy even if
rings are of the same order and having the same number of
zero-divisors (the two 16/12 cases), but, on the other hand, also
a nice hierarchic built-up principle at work: the condensate of
$\widetilde{P}(GF(q) \times R_{\diamondsuit})$  is obviously
isomorphic to $P(GF(2) \times GF(q))$, for $q$ being any power of
a prime. Another noteworthy fact is that, with the exception of
the 16/12B case, all the condensates enjoy the property of having
maximum sets of pairwise distant points of cardinality three, like
the unimodular parts of the parent lines.

Imagine now that the order and complexity of the underlying ring of our
``amphibian" (that is, featuring both the sectors) projective line is so big
that it contains a large number of subrings of various orders such that the
projective lines defined over them are also ``amphibians." This means that our
parent space-time will encompass a unique aggregate of ``sub-space-times" of
different smaller orders and complexity, linked to each other in a particularly
hierarchic way that reflects the relation between the individual subrings. As
we deal with finite rings, these aggregates will always have a limited number
of members. It is likely that for some orders such space-time collections will
be much richer than for others. Hence, our hierarchy will basically be
two-fold: it will run not only within a given order (local), but through
different orders as well (global), the two meeting in the not further reducible
building block --- our above-described ternionic space-time. And such hierarchy
applies not only to the lines (space-times) as a whole, but also separately to
their two sectors (space and time) and to their condensates (imprints of time's
arrows) as well. Couldn't, then, our universe simply be a projective ring line
of a huge, yet still finite order (Table 3), unjustly neglected and
inadequately hidden under a variety of disguises like a differentiable
pseudo-Riemannian manifold, a world of strings and branes, etc.? It would
certainly be desirable to compare the mathematical language of our model with
that of the two currently mainstream competing quantum gravity formalisms,
namely string M-theory and loop quantum gravity (for a most recent relevant
paper in this respect, see Bojowald, 2009); yet, as our familiarity with these
theories is rather superficial and, therefore, insufficient to do such task
properly, we leave it to the reader, if curious enough.

\begin{table}[h]
\begin{center}
\caption{Space-time viewed as the projective line over a
(yet-to-be-found) large finite ring.} \vspace*{0.2cm}
\begin{tabular}{ll}
\hline \hline
Space-time & Projective Ring Line of a Very Large Order\\
Space      &                          Set of Unimodular Points\\
Time   &                              Set of Non-Unimodular Points\\
3D of Space      &               Three Unique Maximum Sets of Mutually Neighbour Unimodulars\\
1D of Time      &                Non-Unimodulars Form One Maximum Set of Mutually Neighbours\\
Arrow of Time     &             Condensation Phenomenon \\
\hline \hline
\end{tabular}
\end{center}
\end{table}

\section{A Few Final Musings}

\vspace*{-.3cm} Apart from the quantum gravity issues, the
above-described projective ring line concept of space-time may also be
relevant and taken as a starting conceptual point for addressing the
behaviour of highly complex (hierarchic) systems. A usual approach
starts with modelling these systems as isolated (closed) and
subject to unitary evolution, i.\,e., symmetric to time reversal.
Any emergence of the ``arrow" of time is ascribed to making these
systems open, that is, coupled to their surrounding. And all kinds
of classical models for open systems are based on the assumption
that this coupling is weak; if the coupling becomes stronger and
stronger we run into serious trouble due to the requirement of the
continuity of both time and space dimensions. In our approach such
problems should not emerge, as at the deepest levels the hierarchy
and complexity principles pertain not only the systems and their
environment(s), but --- as shown in the preceding section --- also
to their ``background" space-times. A closely related question is the fine
structure of hierarchical systems, in particular emergence of qualitative new
kinds of coupling/bonds between subsystems. In our model, this is surmised to be accounted
for by a distinguished role played by rings (and so lines and associated space-times)
of certain orders, especially by power-of-two ones. Here, one can expect a number
of interesting links to various long-standing open mathematical problems
(outer automorphisms of groups, (non-)existence of projective planes of composite orders,
distribution of primes, the Riemann hypothesis, etc.) to emerge in various contexts.

\subsection*{Acknowledgements} The work was partially supported by the VEGA
grant agency projects Nos. 6070 and 7012 (Slovak Republic) and completed within
the framework of the Cooperation Group ``Finite Projective Ring Geometries: An
Intriguing Emerging Link Between Quantum Information Theory, Black-Hole
Physics, and Chemistry of Coupling" at the Center for Interdisciplinary
Research (ZiF), University of Bielefeld, Germany. We thank Mark Stuckey
(Elizabethtown College, PA) for a careful proofreading of the paper.

\subsection*{References}
\begin{list}{}{\labelsep0pt\labelwidth0pt\leftmargin0pt\itemsep1pt}

\item Blunck, A., and Havlicek, H.: 2000, Projective Representations I:
    Projective Lines over Rings, Abh. Math. Sem. Univ. Hamburg, Vol. 70,
    pp. 287--299.
\item Blunck, A., and Herzer, A.: 2005, Kettengeometrien -- Eine
    Einf\"{u}hrung, Berichte aus der Mathematik, Shaker Verlag, Aachen.
\item Bojowald, M.: 2009, A Momentous Arrow of Time, arXiv:0910.3200.
\item Brehm, U., Greferath, and Schmidt, S. E.: 1995, Projective
    Geometry on Modular Lattices, in Handbook of Incidence Geometry, F.
    Buekenhout
      (ed.), Elsevier, Amsterdam, pp. 1115--1142.
\item Havlicek, H., and Saniga, M.: 2008a, Projective Ring Line of an
    Arbitrary Single Qudit, Journal of Physics A: Mathematical and
    Theoretical, Vol. 41, No. 1, 015302, 12pp (arXiv:0710.0941).
\item Havlicek, H., and Saniga, M.: 2008b, Vectors, Cyclic Submodules and
    Projective Spaces Linked with Ternions, Journal of Geometry, Vol. 92,
    pp. 79--90 (arXiv:0806.3153).
\item Herzer, A.: 1995, Chain Geometries, in Handbook of Incidence
    Geometry, F. Buekenhout (ed.),
      Elsevier, Amsterdam, pp. 781--842.
\item Planat, M., and Baboin, A.-C.: 2007, Qudits of Composite Dimension,
    Mutually Unbiased Bases and Projective Ring Geometry, Journal of
    Physics A: Mathematical and Theoretical, Vol. 40, No. 46, pp.
    F1005--F1012 (arXiv:0709.2623).
\item Saniga, M.: 2008, The (Amazing) Architecture of Finite Projective
    Ring Lines, a seminar talk presented at the Institute of Discrete
    Mathematics and Geometry, Vienna University of Technology, Vienna
    (Austria), on March 12, 2008; slides available from
    http://www.ta3.sk/$\widetilde{~~}$msaniga/pub/ftp/- vut$\underline{~}$08.pdf.
\item Saniga, M., Planat, M., and Pracna, P.: 2006, A Classification of
    the Projective Lines over Small Rings II. Non-Commutative Case,
    arXiv:math.AG/0606500.
\item Saniga, M., Planat, M., and Pracna, P.: 2008, Projective Ring Line
    Encompassing Two-Qubits, Theoretical and Mathematical Physics, Vol.
    155, No. 3, pp. 905--913 (arXiv:quant-ph/0611063).
\item Saniga, M., Havlicek, H., Planat, M., and Pracna, P.: 2008, Twin
    "Fano-Snowflakes" over the Smallest Ring of Ternions, Symmetry,
    Integrability and Geometry: Methods and Applications, Vol. 4, Paper
    050, 7 pages (arXiv:0803.4436).
\item Saniga, M., Planat, M., Kibler, M. R., and Pracna, P.: 2007, A
    Classification of the Projective Lines over Small Rings, Chaos,
    Solitons and Fractals, Vol. 33, No. 4, pp. 1095-1102
    (arXiv:math.AG/0605301).
\item Veldkamp, F. D.: 1995, Geometry over Rings, in Handbook of
    Incidence Geometry, F. Buekenhout
      (ed.), Elsevier, Amsterdam, pp. 1033--1084.
\end{list}

\end{document}